\documentclass[aps,pra,floatfix,superscriptaddress,twocolumn,showpacs,10pt]{revtex4-1}
\usepackage{amssymb, amsmath,color,mciteplus,graphicx,subfigure}
\usepackage{calc,epsfig,epstopdf,color,mciteplus,bm,mathrsfs}
\usepackage{times}
\usepackage{float}
\usepackage{multirow}
\usepackage{lipsum}

\begin{document}

\title{State-independent Nonadiabatic Geometric Quantum Gates}

\author{Yan Liang}\email{These two authors contributed equally to this work.}
\affiliation{Guangdong Provincial Key Laboratory of Quantum Engineering and Quantum Materials,
and School of Physics\\ and Telecommunication Engineering, South China Normal University, Guangzhou 510006, China}

\author{Pu Shen}\email{These two authors contributed equally to this work.}
\affiliation{Guangdong Provincial Key Laboratory of Quantum Engineering and Quantum Materials,
and School of Physics\\ and Telecommunication Engineering, South China Normal University, Guangzhou 510006, China}

\author{Li-Na Ji}
\affiliation{Guangdong Provincial Key Laboratory of Quantum Engineering and Quantum Materials,
and School of Physics\\ and Telecommunication Engineering, South China Normal University, Guangzhou 510006, China}

\author{Zheng-Yuan Xue}\email{zyxue83@163.com}
\affiliation{Guangdong Provincial Key Laboratory of Quantum Engineering and Quantum Materials,
and School of Physics\\ and Telecommunication Engineering, South China Normal University, Guangzhou 510006, China}

\affiliation{Guangdong-Hong Kong Joint Laboratory of Quantum Matter,  and Frontier Research Institute for Physics,\\ South China Normal University, Guangzhou 510006, China}

\date{\today}

\begin{abstract}

Quantum computation has demonstrated advantages over classical computation for special hard prolems, where a set of universal quantum gates is essential. Geometric phases, which have built-in resilience
to local noise, have been used to construct quantum gates with excellent performance. However, this
advantage has been smeared in previous schemes. Here, we propose a state-independent nonadiabatic geometric quantum-gate scheme that is able to realize a more fully geometric gate than previous approaches,
allowing for the cancelation of dynamical phases accumulated by an arbitrary state. Numerical simulations demonstrate that our scheme has significantly stronger gate robustness than the previous geometric
and dynamical ones. Meanwhile, we give a detailed physical implementation of our scheme with the
Rydberg atom system based on the Rydberg blockade effect, specifically for multiqubit control-phase
gates, which exceeds the fault-tolerance threshold of multiqubit quantum gates within the considered error range. Therefore, our scheme provides a promising way for fault-tolerant quantum computation in atomic systems.
\end{abstract}

%%%%%%%%%%%%%

\maketitle

\section{Introduction}

As a recently emerged computation pattern based on quantum mechanics \cite{Nielson}, quantum computation has powerful parallel computing capabilities that enable it to exceed classical computation in principle and provide potential solutions to hard computation problems, such as quantum chemistry \cite{SMcArdle}, quantum many-body physics \cite{DYang}, and quantum machine learning \cite{VSaggio}. Moreover, the computational power of a quantum computer increases exponentially with the increase of the qubit number. It is well known that operating qubits to obtain a universal set of quantum gates is the building block for large-scale quantum computation in a fault-tolerant way. However, due to the inevitable noise and the decoherence effect, the physical implementation of quantum computation remains a great challenge. Therefore, realizing quantum gates with high fidelity and strong robustness is essential, especially for two-qubit gates.

Benefiting from global properties, geometric phases can naturally combat certain local noise, and geometric quantum computation (GQC) utilizing either Abelian or non-Abelian holonomy is considered to be an effective way to improve the robustness of quantum gates \cite{Berry1984, FWilczek1984, SLZhu2005, JTThomas2011, PSolinas2012, MJohansson2012}. A GQC based on the Abelian phase is relatively straightforward experimentally, as it only needs the operation of nondegenerate two-level quantum systems to realize quantum gates. However, early GQC schemes were based on adiabatic evolution \cite{PZanardi1999,JAJones2000,GFalci2000,LMDuan2001}, which required longer operating times and imposed additional constraints that introduced more noise and decoherence. To relax the constraints of adiabatic evolution, nonadiabatic GQC (NGQC) was proposed naturally \cite{XBWang2001,SLZhu2002}. Owing to the combination of fast manipulation and strong robustness, NGQC has been developed rapidly \cite{SLZhu2003,XDZhang2005,PZZhao2017,TChen2018,TChen2020,ssl2020,KZLi2020,LNJi2021,CYDing2021,CYDing20212}, and has been experimentally demonstrated in various quantum systems \cite{DLeibfried2003,du2006,YXu2020,PZZhao2021}.

Achieving NGQC requires the elimination of the accompanying dynamical phases. This can be done by setting the dynamical phase to zero at all times, by driving two auxiliary basis vectors to cyclically evolve along the geodesic paths on the Bloch sphere, such as the orange-slice-shaped geometric path \cite{JTThomas2011}. Another approach is to allow the existence of the dynamical phase during the evolution process, but then to set the accumulated dynamical phase at the final time to zero \cite{XDZhang2005,ssl2020,LNJi2021}. However, in all these schemes, an arbitrary superposition state of two auxiliary basis vectors still holds a nonzero dynamical phase \cite{SLZhu2002,SLZhu2003,XDZhang2005}. Additionally, the gate robustness of these schemes is only second order, which is the same as that of the dynamical scheme, in the presence of systematic errors. In particular, when the rotation angle is $\pi$, the geometric rotating gate has the same robustness against systematic errors as the dynamical scheme  \cite{JTThomas2011,TChen2020,PZZhao2021}. Therefore, it is worth exploring whether there exists a more rigorous geometric quantum gate scheme that eliminates the dynamical phase accumulated by an arbitrary initial state, which could lead to significantly stronger gate robustness compared with previous NGQC and dynamical schemes.
%%%%%%%%%%%%%%%%5

%%%%%%%%%%%%%%%%%%
Here, we propose a state-independent NGQC (SINGQC) scheme that not only eliminates dynamical phases accumulated by two auxiliary basis vectors, but also eliminates dynamical phases accumulated by any state. Numerical results show that the gate robustness of our scheme is significantly stronger than both conventional dynamical gate (DG) and previous NGQC schemes. In addition, we present a physical implementation of our scheme in the Rydberg atom system, based on the Rydberg blockade effect. From numerical results, our control phase ($CZ$) gates are exceptionally robust to systematic errors and possess strong immunity to the Rydberg state lifetime, making them more robust than the DG scheme. Without optimization, the fidelities of our SINGQC multi-qubit gates still exceed $99\%$ within the considered error range, representing a major advancement for atomic multi-qubit gates. Overall, our scheme provides an alternative for realizing fault-tolerant quantum computing in atomic systems.

%%%%%%%%%%%

%%%%%%%%%%%%%%

\section{State-independent Geometric Gates}
In this section, we first derive the SINGQC condition realizing the state-independent geometric quantum gates. Then, we design special evolution paths satisfying the SINGQC condition to construct single-qubit gates. Finally, the robustness of our scheme is discussed and compared with the previous single-loop NGQC (SLNGQC) scheme \cite{JTThomas2011,PZZhao2017,TChen2018} and the DG scheme driven by simple resonant pulses (see Appendix A and B for details).

\subsection{The SINGQC condition}
We first proceed to the condition for realizing the state-independent geometric quantum gates (SINGQG), using the reverse engineering of the target Hamiltonian \cite{kang2016,Odelin2019}. For a two-level system, a set of orthogonal auxiliary vectors can be
 \begin{eqnarray}
\label{6}
|\mu_{1}(t)\rangle=&\cos&\frac{\theta(t)}{2}|0\rangle +\sin\frac{\theta(t)}{2}e^{i\varphi(t)}|1\rangle, \notag\\
|\mu_{2}(t)\rangle=&\sin&\frac{\theta(t)}{2}e^{-i\varphi(t)}|0\rangle-\cos\frac{\theta(t)}{2}|1\rangle,
\end{eqnarray}
where $\theta(t)$ and $\varphi(t)$ are the time-dependent parameters. Thus, there is a set of corresponding evolution states $|\psi_{k}(t)\rangle$ ($k=1,2$) satisfying the Schr\"{o}dinger equation $i|\dot{\psi}_{k}(t)\rangle=H(t)|\psi_{k}(t)\rangle$, where $|\psi_{k}(t)\rangle=e^{i\gamma_{k}(t)}|\mu_{k}(t)\rangle$ with $\gamma_{k}(0)=0$ and $\gamma_{k}(t)$ being the accumulated total phase. We assume that the quantum system is controlled by the following Hamiltonian \cite{KZLi2020}
\begin{eqnarray}
\label{3}
H(t)&=&\!\!i\!\sum^2_{k\neq l}\langle\mu_{l}(t)|\dot{\mu}_{k}(t)\rangle|\mu_l(t)\rangle\langle\mu_{k}(t)| \notag \\
    &=&\!\Delta(t) %(|0\rangle\langle0|-|1\rangle\langle1|)
    \sigma_z +[\Omega(t)|0\rangle\langle1|+\rm{H.c.}],
\end{eqnarray}
where $\Delta(t)$ $=$ $\frac{1}{2}\sin^2\theta(t)\dot{\varphi}(t)$ and $\Omega(t)$ $=$ $-i\frac{1}{2}e^{-i\varphi(t)} [\dot{\theta}(t)$ $-i\sin\theta(t)\cos\theta(t)\dot{\varphi}(t)]$. It is easy to verify that the evolution states $|\psi_{k}(t)\rangle$ only accumulate geometric phase, i.e., $\gamma_{k}(t)=i\int_{0}^{t}\langle\mu_{k}(t')|\dot{\mu}_{k}(t')\rangle dt'$. When the auxiliary vectors meet cyclic evolution at the final time $\tau$, i.e., $|\mu_k(\tau)\rangle=|\mu_k(0)\rangle=|\psi_{k}(0)\rangle$, the corresponding evolution operator is
\begin{eqnarray}
\label{4}
U(\tau) = \!\!\!\sum^2_{k=1}|\psi_{k}(\tau)\rangle\langle \psi_{k}(0)| %\notag \\
        =\!\!\!\sum^2_{k=1}e^{i\gamma_{k}(\tau)}|\mu_k(0)\rangle\langle\mu_{k}(0)|,
\end{eqnarray}
where $\gamma_1(\tau)=-\gamma_2(\tau)= -\frac{1}{2}\int_0^{\tau}[1-\cos\theta(t)]\dot{\varphi}(t)dt$. By setting $\theta_0$ = $\theta(0)$, $\varphi_0$ = $\varphi(0)$ and $\gamma=\gamma_1(\tau)$, we further obtain the evolution operator in the computation space spanned by $\{|0\rangle,|1\rangle\}$ as
\begin{eqnarray}
\label{8}
U(\tau)=e^{i\gamma \mathbf{n} \cdot \mathbf{\sigma}},
\end{eqnarray}
where $\mathbf{n}=(\sin\theta_0\cos\varphi_0, \sin\theta_0\sin\varphi_0, \cos\theta_0)$ is a unit vector and $\mathbf{\sigma}=(\sigma_x, \sigma_y, \sigma_z)$ is a vector of standard Pauli operators. Obviously, the evolution operation $U(\tau)$ is an arbitrary rotation gate around the rotation axis $\mathbf{n}$ by the rotation angle $-2\gamma$. %Meanwhile, it is the arbitrary single-qubit nonadiabatic geometric quantum gates, because $\gamma$ is the pure geometric phase.
However, considering an arbitrary initial state %$|\Psi(0)\rangle$, which can be written as
$|\Psi(0)\rangle=C_1|\psi_{1}(0)\rangle+C_2|\psi_{2}(0)\rangle$, where $C_1$ and $C_2$ are the nonzero complex numbers that satisfy $|C_1|^2+|C_2|^2=1$, we find that, during the evolution time $[0, \tau]$, the accumulated dynamical phase  $\gamma_d(\tau)$ % by $|\Psi(t)\rangle=C_1|\psi_{1}(t)\rangle+C_2|\psi_{2}(t)\rangle$
is no longer zero,  but
 \begin{eqnarray}
\label{5}
\gamma_d(\tau) &=&\int_{0}^{\tau} \langle\Psi (t)|H(t)|\Psi (t)\rangle dt \notag \\
     &=& \int_{0}^{\tau} \{ iC^{*}_1C_2 e^{i[\gamma_2(t)-\gamma_1(t)]}\langle\mu_{1}(t)|\dot{\mu}_{2}(t)\rangle \notag \\
       && +iC^{*}_2C_1 e^{i[\gamma_1(t)-\gamma_2(t)]}\langle\mu_{2}(t)|\dot{\mu}_{1}(t)\rangle \}dt.
\end{eqnarray} 
To obtain $\gamma_d(\tau)=0$, which is equivalent to  $\int_{0}^{\tau} i{\rm exp}\{i[\gamma_2(t)-\gamma_1(t)]\} \langle\mu_{1}(t)|\dot{\mu}_{2}(t)\rangle  dt=0$,
the following SINGQC condition should be met:
 \begin{eqnarray}
\label{10}
 \int_0^{\tau}e^{i\int_0^{t}[1-\cos\theta(t')]\dot{\varphi}(t')dt'}e^{-i\varphi(t)}[i\dot{\theta}(t)+\sin\theta(t)\dot{\varphi}(t)]dt
 =0.  \notag \\
\end{eqnarray} 
%We call Eq. (\ref{10}) as the SINGQC condition.

\begin{figure}[tb]
  \centering
  \includegraphics[width=0.9\linewidth]{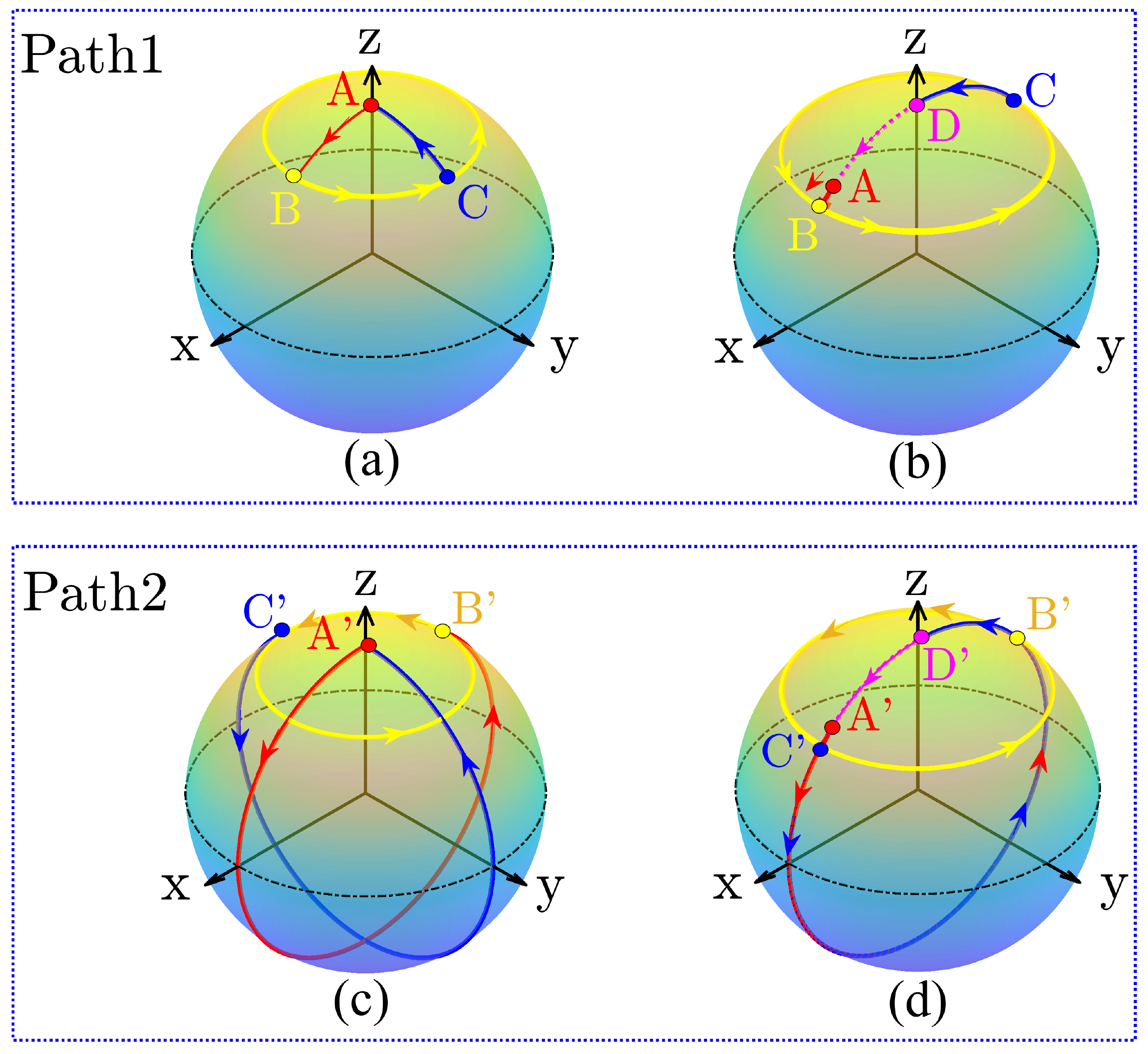}
\caption{The evolution path of $|\psi_{1}(t)\rangle$ on the Bloch sphere for different gates, where (a) and (b) belong to Path~1 corresponding to the condition in Eq. (\ref{path1}), and (c) and (d) belong to Path~2 corresponding to the condition in Eq. (\ref{path2}). (a) Path~1 for the $S$ gate. (b) Path~1 for the $H$ gate. (c) Path~2 for the $S$ gate. (d) Path~2 for the $H$ gate. 
 The evolution process can be described using spherical coordinates $[\Theta(t),\Phi(t)]$. Start from point $A$  ($A'$) $(\theta_0,0)$ of Path~1 (Path~2). First, rotate $(\theta_1-\theta_0)$ [$(2\pi-\theta_1-\theta_0)$] counterclockwise around the $y$-axis to point $B$ $(\theta_1,0)$ [$B'$ $(\theta_1,\pi)$ ]. Then, rotate $2\pi/\cos\theta_1$ counterclockwise around the $z$-axis to reach point $C$ $(\theta_1,2\pi/\cos\theta_1)$ [$C'$ $(\theta_1,\pi+2\pi/\cos\theta_1$]. Then, rotate $(\theta_1-0)$ [($2\pi-\theta_1$)] clockwise around the axis of \textbf{n}= $[\cos(\pi/2+2\pi/\cos\theta_1),\sin(\pi/2+2\pi/\cos\theta_1),0]$ to piont $D(0,0)$ [$D'(0,0)$]. Finally, return to the initial point $A(\theta_0,0)$ [$A'(\theta_0,0)$] by rotating $(\theta_0-0)$ counterclockwise around the $y$-axis. Note that the fourth segment is unnecessary for the $S$ gate since $\theta_0=0$.} \label{fig1}
\end{figure}

\subsection{Arbitrary single-qubit gate}

In fact, to realize the SINGQC scheme, there are many ways to satisfy Eq. (\ref{10}). Here, we design special evolution paths satisfying the SINGQC condition to construct arbitrary single-qubit gates. Without loss of generality, we divide the evolution process into four segments, and the Hamiltonian from Eq. (\ref{3}) in each segment is
\begin{equation}
\label{11}
\begin{cases}
    H_1(t) =\Omega_1e^{-i(\frac{\pi}{2}+\varphi_0)}|0\rangle\langle 1|+\rm{H.c.},\qquad \quad \ \ \ t\in[0,\tau_1], \\

    H_2(t) =\left[\Omega_2e^{-i\varphi(t)}|0\rangle\langle 1|+\rm{H.c.}\right]+\Delta\sigma_z,  \ \      t\in(\tau_1,\tau_2], \\
    H_3(t) =\Omega_3e^{-i[\frac{\pi}{2}+\varphi(\tau_2)]}|0\rangle\langle 1|+\rm{H.c.},\quad \quad \ \ \ t\in(\tau_2,\tau_3],\\
    H_4(t) =\Omega_4e^{-i(\frac{\pi}{2}+\varphi_0)}|0\rangle\langle 1|+\rm{H.c.},\quad \quad \ \ \ \ \ \ t\in[\tau_3,\tau],\\
\end{cases}
\end{equation}
and the requirements of the parameters are
\begin{equation}
\label{path1}
\begin{cases}
    \int_0^{\tau_1}\Omega_1dt=\int_0^{\tau_1}\frac{\dot{\theta}(t)}{2}dt=\frac{\theta_1-\theta_0}{2}, \\
    \Omega_2=-\frac{1}{2}\sin\theta_1\cos\theta_1\dot{\varphi}(t); \ \Delta=\frac{1}{2}\sin^2\theta_1 \dot{\varphi}(t), \\
    \int_{\tau_2}^{\tau_3}\Omega_3dt=\int_{\tau_2}^{\tau_3}\frac{\dot{\theta}(t)}{2}dt=\frac{0-\theta_1}{2}, \\
     \int_{\tau_3}^{\tau}\Omega_4dt=\int_{\tau_3}^{\tau}\frac{\dot{\theta}(t)}{2}dt=\frac{\theta_0-0}{2},
\end{cases}
\end{equation}
with $\theta_1=\theta(\tau_1)$. In the implementation of the Hamiltonian, the variable $\varphi(t)$ is set to be time dependent in the second segment and time independent in the other three segments, while $\theta(t)$ is set to be time independent in the second segment and time dependent in the other three segments. Under these settings, Eq. (\ref{10}) reduces to
\begin{eqnarray}
\label{13}
 \int_{\tau_1}^{\tau_2}e^{-i\int_0^{t}\cos\theta_1\dot{\varphi}(t)dt'} \sin\theta_1\dot{\varphi}(t) dt=0.
\end{eqnarray}
%the accumulated dynamical phase during the whole evolution time $[0, \tau]$ will be removed, i.e., $\gamma_d(\tau)=0$ by Eqs. (\ref{5}) and (\ref{10}).
Thus, to satisfy Eq. (\ref{13}), we choose $\varphi(t)=\frac{2\pi(t-\tau_1)}{\cos\theta_1(\tau_2-\tau_1)}+\varphi_0$ $(t\in(\tau_1,\tau_2])$ in the second segment,  and the modulation of this time-dependent phase can be realized with a phase modulator \cite{Todd2015,Melikyan2014,Figgatt2019}. Meanwhile, we can obtain the specific expression of the half rotation angle $\gamma$ in the evolution operator, i.e.,
 \begin{eqnarray}
\label{theta1}
 \gamma&=& \frac{1}{2}\int_0^{\tau}[1-\cos\theta(t)]\dot{\varphi}(t)dt \notag \\
       &=&\frac{1}{2}\int_{\tau_1}^{\tau_2}[1-\cos\theta_1]\dot{\varphi}(t)dt \notag \\
       &=&\left(\frac{1}{\cos\theta_1}-1\right)\pi,
\end{eqnarray}
where $\theta_1=\arccos\frac{\pi}{\gamma+\pi}$ depends on the specific $\gamma$. Therefore, after setting gate parameters $(\theta_0,\varphi_0,\gamma)$, we can construct the arbitrary single-qubit gates for the SINGQC scheme by applying the Hamiltonian in Eq. (\ref{11}). It is worth noting that, under the control of the Hamiltonian in Eq. (\ref{11}), $|\psi_1(t)\rangle$ and $|\psi_2(t)\rangle$ from an arbitrary evolution state $|\Psi(t)\rangle$ will cyclically evolve on Bloch sphere, where $|\Psi(t)\rangle=C_1|\psi_{1}(t)\rangle+C_2|\psi_{2}(t)\rangle$.
%%%

%%
To illustrate the evolution process more clearly, we use the $S$ and Hadamard ($H$) gates as examples, with $(\theta_0,\varphi_0,\gamma)$ set to $(0,0,\pi/4)$ and $(\pi/4,0,\pi/2)$, respectively. We plot the evolution trajectories (Path~1) of $|\psi_1(t)\rangle$ on the Bloch sphere, as shown in Figs. \ref{fig1}(a) and (b) for the $S$ and $H$ gates, respectively.
From the perspective of $|\psi_1(t)\rangle$, the evolution process can be described as follows. First, $|\psi_1(t)\rangle$ evolves from point $A$ to $B$ under the control of Hamiltonian $H_1(t)$ in Eq. (\ref{11}) for a time duration of $[0,\tau_1]$ with $\tau_1=|(\theta_1-\theta_0)/(2\Omega_1)|$. Then, from point $B$, the system Hamiltonian is switched to $H_2(t)$, and $|\psi_1(t)\rangle$ reaches point $C$ after a time duration of $\tau_2-\tau_1=|\pi\sin\theta_1/\Omega_2|$. Next, from point $C$, the quantum system is governed by the Hamiltonian $H_3(t)$ and $|\psi_1(t)\rangle$ reaches point $D$ after a time duration of $\tau_3-\tau_2=|\theta_1/(2\Omega_3)|$. Finally, $|\psi_1(t)\rangle$ will be back to the starting point $A$ by applying a control field $\Omega_4$ for a duration of $[\tau_3,\tau]$ with $\tau-\tau_3=|\theta_0/(2\Omega_4)|$, where the corresponding Hamiltonian is $H_4(t)$ in Eq. (\ref{11}). It is worth noting that the fourth segment is not necessary for the $S$ gate since $\theta_0=0$, resulting in a duration of $\tau-\tau_3=|\theta_0/(2\Omega_4)|=0$.

%From this perspective, the evolution process can be described with spherical coordinate $[\Theta(t),\Phi(t)]$ as: Starting from point A$(\theta_0,0)$, firstly, rotate $(\theta_1-\theta_0)$ counterclockwise around the y-axis to B$(\theta_1,0)$; Secondly, rotate $\frac{2\pi}{\cos\theta_1}$ counterclockwise around the z-axis to C$(\theta_1,\frac{2\pi}{\cos\theta_1}-2\pi)$; Thirdly, rotate $(\theta_1-0)$ clockwise around the axis of \textbf{n}= $[\cos(\frac{\pi}{2}+\frac{2\pi}{\cos\theta_1}-2\pi),\sin(\frac{\pi}{2}+\frac{2\pi}{\cos\theta_1}-2\pi),0]$ to D$(0,0)$, and finally return to the initial point A$(\theta_0,0)$ by rotating $(\theta_0-0)$ counterclockwise around the y-axis. Similarly, the fourth segment is not necessary for the $S$ and $T$ gates since $\theta_0=0$.

\begin{figure}[t]
  \centering
  \includegraphics[width=\linewidth]{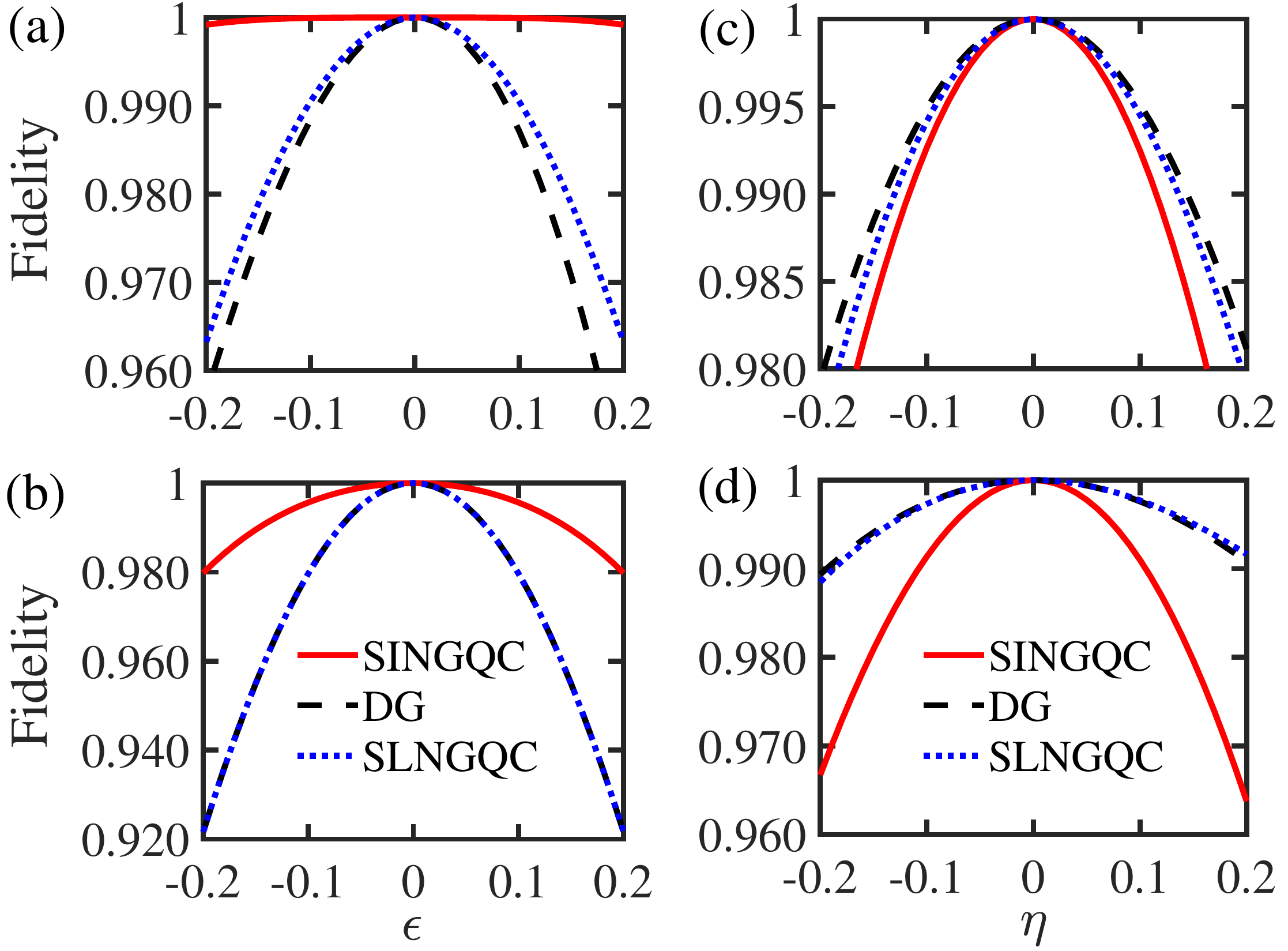}
\caption{ Gate fidelities evolving along Path~1 as functions of the control error without decoherence, for the results of the (a) $S$ and (b) $H$ gates. Gate fidelities evolving along Path~1 as functions of detuning error without decoherence, with the results of the $S$ and $H$ gates being shown in (c) and (d), respectively.
%The solid-red, dashed-black, and dotted-blue lines denote the results of the SINGQC, DG, and SLNGQC schemes, respectively.
}\label{fig2}
\end{figure}

\subsection{Gate performance}

We turn to test the performance of the implemented SINGQC  gates and compare it with the previous schemes.
The performance of quantum gates in an open quantum system can be evaluated by the Lindblad master equation
\begin{eqnarray}
\label{EqMaster}
\dot\rho&=&-i[H'(t), \rho]+\frac {1} {2}\sum_{j=-,z}\Gamma_{j}L(\sigma_{j}),
\end{eqnarray}
where the quantum system is controlled by $$H'(t)=(1+\epsilon)H(t)+\frac{\eta}{2}\Omega_i\sigma_z$$ with $\epsilon$ and $\eta$ being the error fractions of the control and detuning errors, respectively; $\rho$ is the density matrix of the quantum system; $L(A)=2A\rho A^{\dag}-A^{\dag}A\rho-\rho A^{\dag}A $ is the Lindbladian operator with $\sigma_-=|0\rangle\langle 1|$, $\sigma_z=|1\rangle\langle 1|-|0\rangle\langle0|$; $\Gamma_-$ and $\Gamma_z$ are the decay and dephasing rates, respectively. We also consider the influence of phase error ($\Omega e^{i\phi}\rightarrow \Omega e^{i(1+\chi)\phi}$) in Appendix C.

To show the noise-resilient advantage of our SINGQC scheme, we first consider only the influence of the control error, which destroys the cyclic evolution and introduces a nonzero dynamical phase. The gate fidelity is defined as $F=\frac{1}{6}\sum_{l=1}^6\langle\Psi_l(0)|U(\tau)^{\dag}\rho U(\tau)|\Psi_l(0)\rangle$, where the six initial states $|\Psi_l(0)\rangle$ are $|0\rangle$, $|1\rangle$, $(|0\rangle+|1\rangle)/\sqrt{2}$, $(|0\rangle-|1\rangle)/\sqrt{2}$, $(|0\rangle+i|1\rangle)/\sqrt{2}$ and $(|0\rangle-i|1\rangle)/\sqrt{2}$, respectively. For simplicity, we set $|\Omega_{i}(t)|=\Omega$ to be the time-independent driving amplitude. Figures. 2(a) and 2(b) show plots of the fidelities of the $S$ and $H$ gates as a function of $\epsilon$, respectively. Our scheme exhibits excellent robustness to control error, and the gate fidelity of the $S$ gate remains above $99.9\%$ in the range of the error ratio $\epsilon\in[-0.2, 0.2]$, which far exceeds that of the DG and SLNGQC schemes. The $H$ gate also far outperforms the SLNGQC and DG schemes in terms of robustness to control error and, when the error ratio is $\epsilon=|0.2|$, the gate fidelity of the SINGQC scheme is $6\%$ higher than that of the corresponding SLNGQC and DG schemes. On the other hand, we also plot gate fidelities as a function of the detuning error of the $S$ and $H$  gates, as shown in Figs. 2(c) and 2(d), respectively. However, the results are not satisfactory.

\begin{figure}[tbp]
  \centering
\includegraphics[width=\linewidth]{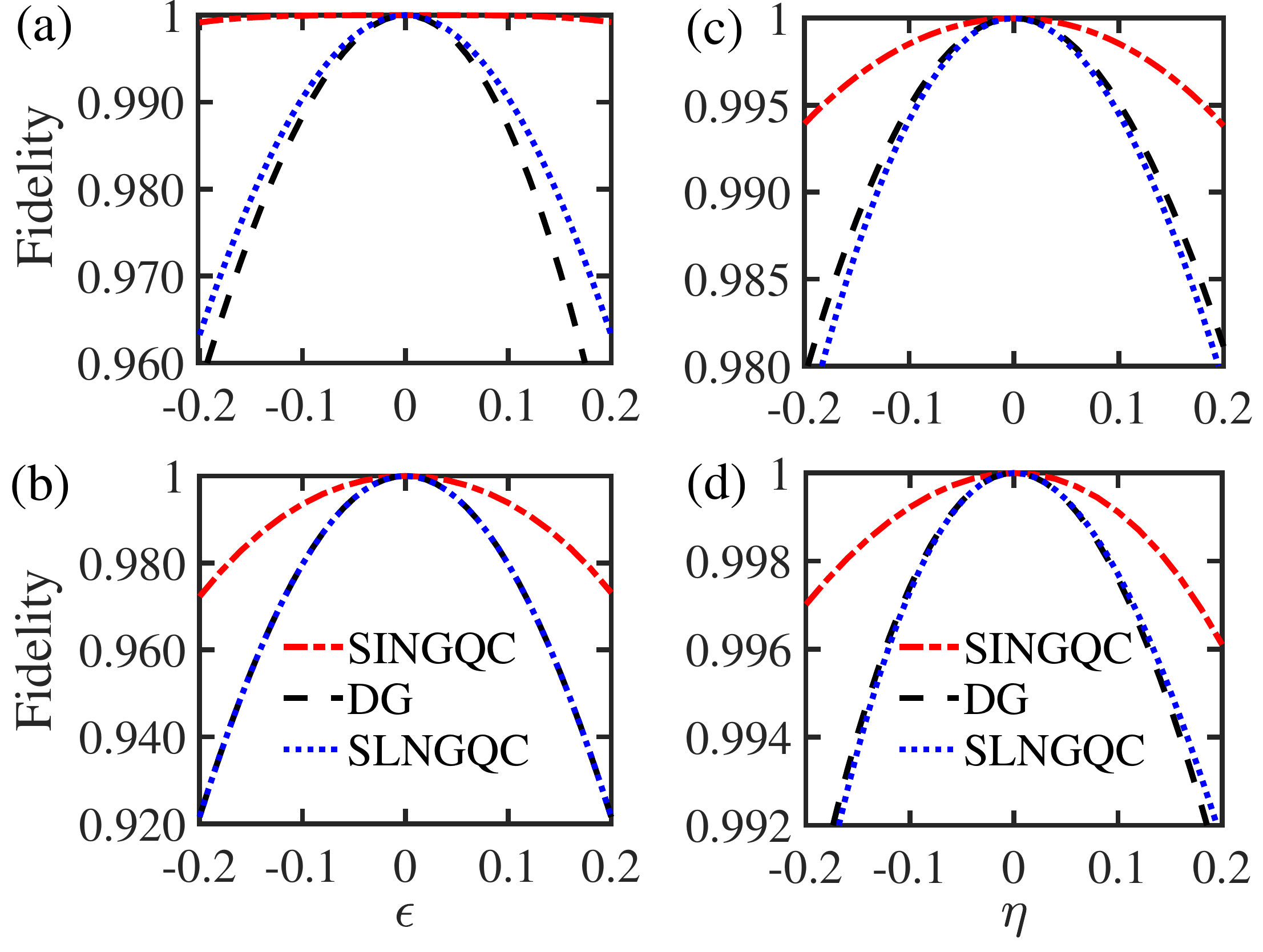}
\caption{ Gate fidelities evolving along Path~2 as functions of the control error without decoherence. The results of the $S$ and $H$ gates are shown in (a) and (b), while their performance against detuning error are shown in (c) and (d), respectively. %The dotted-red, dashed-black, and dotted-blue lines denote the results of the SINGQC, DG, and SLNGQC schemes, respectively.
}\label{fig3}
\end{figure}

\subsection{An alternative scheme and its performance}

In order to make our scheme robust to both control and detuning errors, we modify the requirements for the parameters in Eq. (\ref{path1}) as follows:
\begin{equation}
\label{path2}
\left\{
\begin{aligned}
    \int_0^{\tau_1}\Omega_1dt=\int_0^{\tau_1}\frac{\dot{\theta}(t)}{2}dt=\frac{2\pi-\theta_1-\theta_0}{2},  \qquad \qquad \quad\\
    \Omega_2=\frac{1}{2}\sin\theta_1\cos\theta_1\dot{\varphi}(t); \ \Delta=\frac{1}{2}\sin^2\theta_1\dot{\varphi}(t), \qquad \quad \\
    \int_{\tau_2}^{\tau_3}\Omega_3dt=\int_{\tau_2}^{\tau_3}\frac{\dot{\theta}(t)}{2}dt=\frac{0-(2\pi-\theta_1)}{2}, \qquad \qquad \quad \\
     \int_{\tau_3}^{\tau}\Omega_4dt=\int_{\tau_3}^{\tau}\frac{\dot{\theta}(t)}{2}dt=\frac{\theta_0-0}{2}. \qquad \qquad\qquad\qquad \quad  \\
\end{aligned}
\right.
\end{equation}
Under these settings, the evolution trajectories (Path~2) of $|\psi_1(t)\rangle$ on the Bloch sphere are shown in Figs. 1(c) and 1(d), for the $S$ and $H$ gates, respectively. The evolution process can be described as follows: In the first step, $|\psi_1(t)\rangle$ evolves from point $A'$ to $B'$ by the control of Hamiltonian $H_1(t)$ in Eq. (\ref{11}) for a duration of $[0,\tau_1]$, but with $\tau_1=|(2\pi-\theta_1-\theta_0)/(2\Omega_1)|$.  Subsequently, the system Hamiltonian is changed to $H_2(t)$, and $|\psi_1(t)\rangle$ reaches point $C'$ after a duration of $\tau_2-\tau_1=|\pi\sin\theta_1/\Omega_2|$. Next, $|\psi_1(t)\rangle$ reaches point $D'$ by the control of Hamiltonian $H_3(t)$ for a duration of $\tau_3-\tau_2=|(\theta_1-2\pi)/(2\Omega_3)|$. Finally, the system Hamiltonian is changed to be $H_4(t)$ during $[\tau_3,\tau]$ with $\tau-\tau_3=|\theta_0/(2\Omega_4)|$, and thus $|\psi_1(t)\rangle$ will be back to the starting point $A'$.

\begin{figure}[t]
  \centering
  \includegraphics[width=\linewidth]{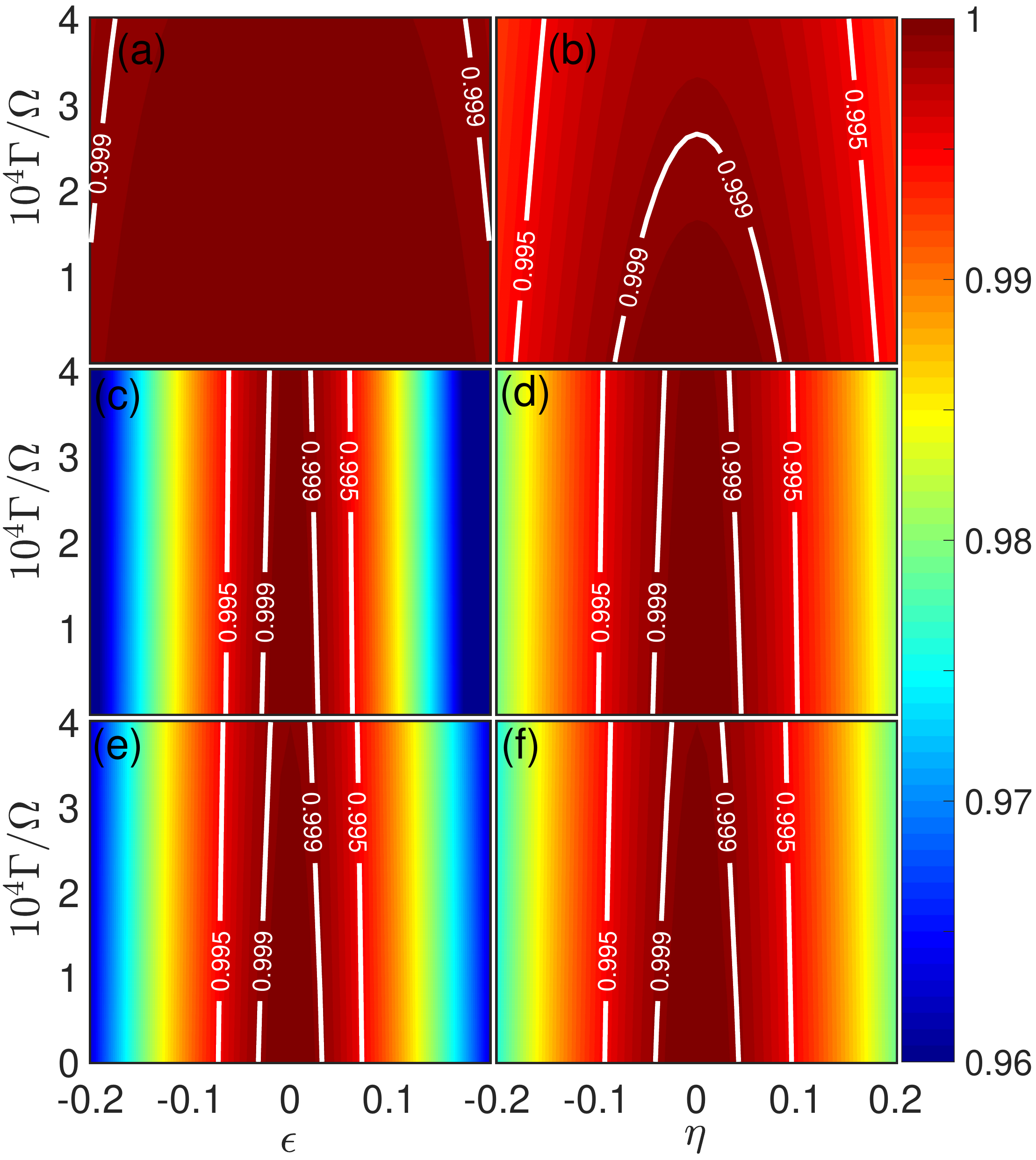}
  \caption{ The robustness of the $S$ gate in the (a) SINGQC along Path~1, (c) DG, and (e) SLNGQC cases, considering both the control error and the decoherence effect. The robustness of the $S$ gate in the (b) SINGQC along Path~2, (d) DG, and (f) SLNGQC cases, considering both the detuning error and the decoherence effect.}\label{fig4}
\end{figure}

The gate robustness against the control error for the $S$ and $H$ gates are shown in Figs. 3(a) and (b), respectively. The results indicate that Path~1 and Path~2 have similar robustness to control error and are far superior to SLNGQC and DG schemes. Furthermore, Path2 also achieves satisfactory results in terms of robustness to detuning errors, with a fidelity higher than $99.3\%$ for all gates across the error rate range of $\eta\in[-0.2,0.2]$, as shown in Figs. 3(c) and (d). Therefore, our scheme can possess stronger robustness to both control and detuning errors than previous schemes. %, which breaks the fact that the robustness of the NGQC schemes cannot be stronger than that of the DG scheme when the rotation angle is $\pi$.

%The evolution process can be described with spherical coordinate $[\Theta(t),\Phi(t)]$ as: Starting from point A $(\theta_0,0)$, firstly rotate $(2\pi-\theta_1-\theta_0)$ degrees counterclockwise around the y-axis to point B $(\theta_1,\pi)$; secondly, rotate $\frac{2\pi}{\cos\theta_1}$ degrees counterclockwise around the Z-axis to point C $(\theta_1,-\pi+\frac{2\pi}{\cos\theta_1})$; thirdly, rotate $(2\pi-\theta_1-0)$ degrees clockwise around the axis of \textbf{n}= $[\cos(-\frac{\pi}{2}+\frac{2\pi}{\cos\theta_1}),\sin(-\frac{\pi}{2}+\frac{2\pi}{\cos\theta_1}),0]$ to point D $(0,0)$, and finally rotate $(\theta_0-0)$ degrees counterclockwise around the y-axis to return to the initial point A $(\theta_0,0)$.

\subsection{Gate performance under decoherence}

However, since the evolution along Path~2 experiences a longer trajectory, it inevitably requires a longer operation time. Therefore, considering the decoherence, the evolution method of Path~1 is more suitable for an actual physical system affected greatly by control error, while the evolution method of Path~2 is more inclined to be selected for a physical system affected greatly by detuning error. Next, taking the $S$ gate as an example, we comprehensively analyze the effects of systematic error and decoherence (with uniform rates $\Gamma$). %The evolution Path 1 and Path 2 are adopted to analyze the influence of control error and detuning error, respectively.
As shown in Fig. \ref{fig4},
%after comprehensively considering the influence of both decoherence and systematic error,
our SINGQC scheme performs best in both control and detuning error. Remarkably, even with a decoherence rate of $\Gamma=4\times 10^{-4}\Omega $, the gate fidelities of our scheme can exceed $99.5\%$ within the control error range of $\epsilon\in[-0.2,0.2]$ and the detuning error range of $\eta\in[-0.15,0.15]$, as shown in Figs. \ref{fig4}(a) and (b), respectively. %Obviously, our scheme is more robust than the SLNGQC and DG schemes.

\section{ Physical realization}
By exciting neutral atoms into the high principal quantum number state, Rydberg atoms have received extensive theoretical and experimental attention \cite{Saffman2010, Jaksch2000,Isenhower2010,Levine2019,CPShen2019,Saffman2016,wujinlei2021,MengLi2021,YanLiang2022} because of their excellent atomic properties. In this section, we propose to implement the SINGQC scheme in the Rydberg atomic system by encoding qubit bases with a pair of long-lived hyperfine ground clock states of typical alkali atoms.

\subsection{Single-qubits quantum gate}
As shown in Fig. \ref{Fig5}(a), quantum information is encoded in two magnetic-field-insensitive hyperfine ground states $|0\rangle\equiv|5S_{1/2}, F=1, m_F=0\rangle$ and $|1\rangle\equiv|5S_{1/2}, F=2,  m_F=0\rangle$, which can be controlled by a two-photon Raman transition \cite{Saffman12010,Saffman2016}. This process can be realized by using a single ground-state Rabi laser with two frequency components generated by current modulation of a diode laser \cite{Saffman12010}. The laser is detuned from the $5P_{3/2}$ excited state by $\delta$. Usually, the Raman lasers are far detuned from the short-lived electronically excited states $5P_{3/2}$, so the decoherence of excited states can be neglected. 
In this case, the Hamiltonian of the single-qubit gate is in the same form as Eq. (\ref{3}),
with $\Omega\approx\Omega_A\Omega_B/\delta$, and $\Delta\approx\omega_A+\omega_B-\omega_0+(\Omega^2_A-\Omega^2_B)/\delta$, where $\Omega_A$, $\Omega_B$ characterize the coupling strengths for the two Raman fields, $\omega_A$, $\omega_B$ are the corresponding coupling field frequencies, $\phi(t)$ is the local phase, and $\omega_0$ is the atomic resonance frequency \cite{MMorgado2021}.
%The Hamiltonian in Eq. (\ref{1qubit-totalH}) has and thus can naturally be used to realize the arbitrary SINGQC single-qubit gates.
%However, off-resonant scattering from the intermediate state results in extra decay (and dephasing) of the atomic states during state manipulations, weakening the stability of the ground state.
%We adopt the following set of experimental parameters \cite{DDYavuz2006}:
When $\Omega=2\pi \times1.36 \ {\rm MHz }$, and $\Gamma=1.15 \ {\rm kHz }\approx\Omega/7400$ \cite{DDYavuz2006}, the gate fidelity can exceed $99.9\%$ within the control error range of $\epsilon\in[-0.2,0.2]$, and over $99.3\%$ within the detuning error range of $\eta\in[-0.2,0.2]$, as shown in Figs. \ref{fig4}(a) and (b), respectively.

\subsection{Multiqubit quantum gate}

In addition to single-qubit quantum gates, the implementation of nontrivial two-qubit gates is crucial for universal quantum computation. While an arbitrary multiqubit quantum gate can be decomposed into several single- and two-qubit quantum gates, it is still worthwhile to directly implement the $N$-qubit ($N>3$) quantum gate because it can reduce the complexity of a large quantum circuit. Rydberg atoms are promising platforms for the implementation of multiqubit quantum gates owing to their excellent interaction properties. In the following, we pay attention to directly realizing SINGQC multiqubit quantum gates, with the $CZ$ gate being one of the typical examples.

\begin{figure}[tbp]
  \centering
  \includegraphics[width=1\linewidth]{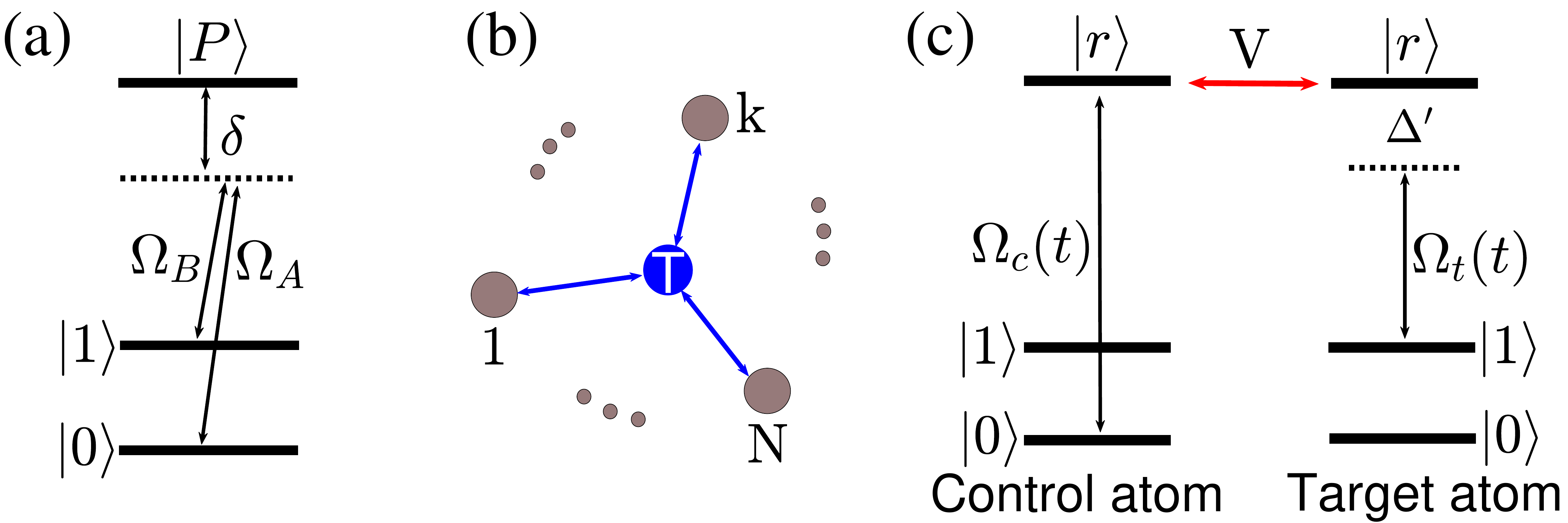}
\caption{ (a) The relevant atomic level for the realization of a single-qubit SINGQC gate. (b) Illustration of a multiqubit $C_{N} Z$ gate, where $N$ represents the number of control atoms and $T$ denotes the target atom. (c) The level diagrams of the ontrol and target atoms.}
\label{Fig5}
\end{figure}

As shown in Fig. \ref{Fig5}(b), we consider $N+1$ Rubidium atoms, where $N$ is the number of control atoms and $T$ denotes the target atom. The related energy levels of each atom are $|0\rangle\equiv|5S_{1/2}, F=1, m_F=0\rangle$, $|1\rangle\equiv|5S_{1/2}, F=2,  m_F=0\rangle$, and $|r\rangle\equiv|83S,\;J=1/2, \; m_J=1/2\rangle$. The quantum information is encoded in two stable ground states $|0\rangle$ and $|1\rangle$, and the Rydberg state $|r\rangle$ is acting as the auxiliary state \cite{MengLi2021,WLi2014}. The states $|0\rangle$ and $|r\rangle$ of the control atoms are coupled resonantly by the Rabi frequency $\Omega_c(t)$,  and states $|1\rangle$ and $|r\rangle$ of the target atom are coupled nonresonantly via the Rabi frequency $\Omega_t(t)$ with a detuning $\Delta'$, as shown in Fig. \ref{Fig5}(c). Rydberg-Rydberg interaction strength $V$ can be adjusted by precise control of atom positions with optical tweezer arrays \cite{HLevine2018,AOmran2019}. The total Hamiltonian of the multiqubit system reads
 \begin{eqnarray}
\label{2qubit-totalH}
 \mathcal{H}_2(t)&=&(1+\epsilon_c)\mathcal{H}_{c}(t)+(1+\epsilon_t)\mathcal{H}_{t}(t)+\mathcal{H}_ V \notag \\
&+&\eta'\Omega'(|r\rangle_c \langle r|+|r\rangle_t \langle r|),
\end{eqnarray}
where $\mathcal{H}_c(t)= \sum_{k=1}^N\Omega_{\rm c}(t)|r\rangle_k\langle 0|+{\rm H.c.} $ is the Hamiltonian of the control atoms with a time-dependent Rabi frequency of $\Omega_c(t)=\bar{\Omega}_c\cos\omega t$ and $\mathcal{H}_t(t)= \Omega_t(t)e^{-i\Delta't} |r\rangle_t\langle 1| +\rm{H.c.} $ is the Hamiltonian of the target qubit. The interaction Hamiltonian between the Rydberg states is represented by
$$\mathcal{H}_ V=\sum_{k>j=1}^N(V_{jk}|rr\rangle_{jk}\langle rr|+V_{kt}|rr\rangle_{kt}\langle rr|),$$
where $V_{jk}$ is the Rydberg-Rydberg interaction between control atoms, and $V_{kt}$ is the Rydberg-Rydberg interaction between the $k\rm{th}$ control atom and the target atom. For the sake of simplicity, we supposethat $V_{jk}=V_c$ and $V_{kt}=V_{t}=\omega$. $\Omega'$ represents the amplitude of the Rabi frequency $\Omega_t(t)$, $\epsilon_c$ ($\epsilon_t$) represents the control error of the control (target) atoms, and $\eta'$ is the detuning error. Under the conditions of a strong Rydberg-Rydberg interactions mechanism $V_{t}\gg\{\bar{\Omega}_c,\Omega'\}$, and when $\bar{\Omega}_{c}\gg\Omega'$, we can derive the effective Hamiltonian of the multiqubit system as \cite{wujinlei2021}
\begin{eqnarray}
\label{multi-effective}
\mathcal{H}_{eff}(t)&=&\left(\otimes_j^N|1\rangle_j\langle1|\right )\otimes \mathcal{H}_{t}(t).
\end{eqnarray}
In the rotation frame of unitary transformation ${\rm \exp}(-iht)$ with $h=\frac{\Delta'}{2}(|1r\rangle_{ct}\langle 1r| - |11\rangle_{ct}\langle 11|)$, the effective Hamiltonian becomes
\begin{eqnarray}
\label{multi-final}
\mathcal{H}'_{eff}(t)&=&[\Omega_t(t)|11\rangle_{ct}\langle 1r|+{\rm H.c.}] \notag \\
&+&\frac{\Delta'}{2}(|11\rangle_{ct}\langle 11|-|1r\rangle_{ct}\langle 1r|)],
\end{eqnarray}
where $|1i\rangle_{ct}=|1\rangle_c\otimes|i\rangle_t$ ($i=1,\ r$) and $|1\rangle_c$ means all control atoms are in the $|1\rangle$ state. It is obvious that Eq. (\ref{multi-final}) with basis vectors $|11\rangle_{ct},\ |1r\rangle_{ct}$ possesses the same form as Eq. (\ref{3}), thereby we can implement a geometric phase only on the computation basis $|11\rangle_{ct}$. %On the other hand, if the control atoms exist the occupations of $|0\rangle$ state, the evolution of the whole system is frozen.
As a result, a multiqubit controlled-phase gate (\textbf{$C_N$}Z) can be obtained, and what is more, the operation time here is independent of the involved number of atoms.

For the realization of the proposed multi-qubit model with Rydberg atoms, we need the atomic arrangement structure as depicted in Fig. 5(b). This structure can be
realized by a defect-free three-dimensional array with the control atoms distributed on the spherical surface \cite{Kumar2018,Barredo2018}. 
 Based on this array, one can greatly increase the available number of control atoms.

\begin{figure}[tbp]
  \centering
  \includegraphics[width=\linewidth]{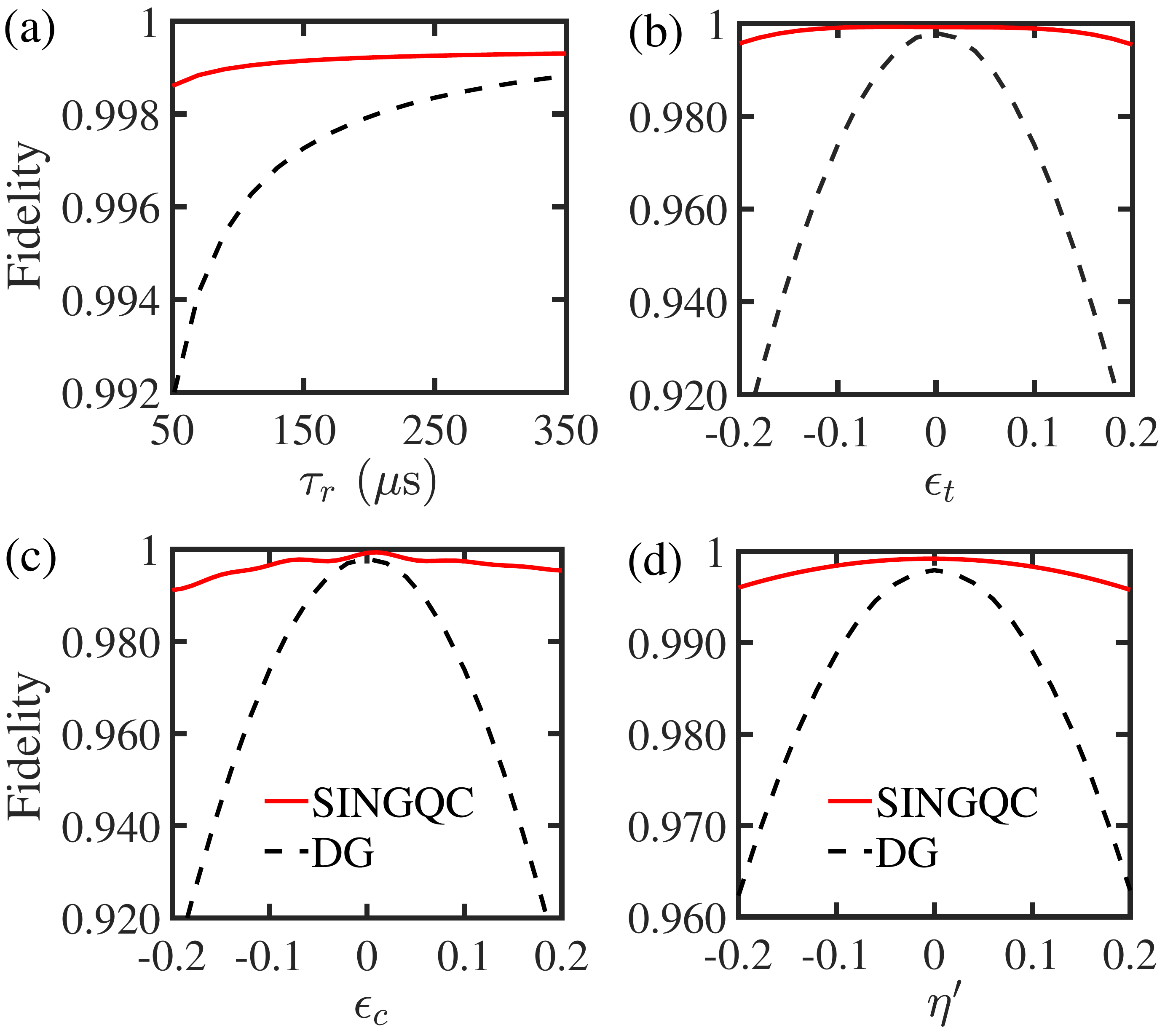}
\caption{ The performance of the $CZ$ gate implemented with the SINGQC scheme (solid red line) and the DG scheme (dashed black line). (a) Fidelities of the $CZ$ gate under different Rydberg state lifetimes. Fidelities of the $CZ$ gate with respect to the (b) control error of the target atom, (c) the control error of the control atoms, and (d) the detuning error. } 
\label{Fig6}
\end{figure}

\begin{figure}[tbp]
  \centering
  \includegraphics[width=\linewidth]{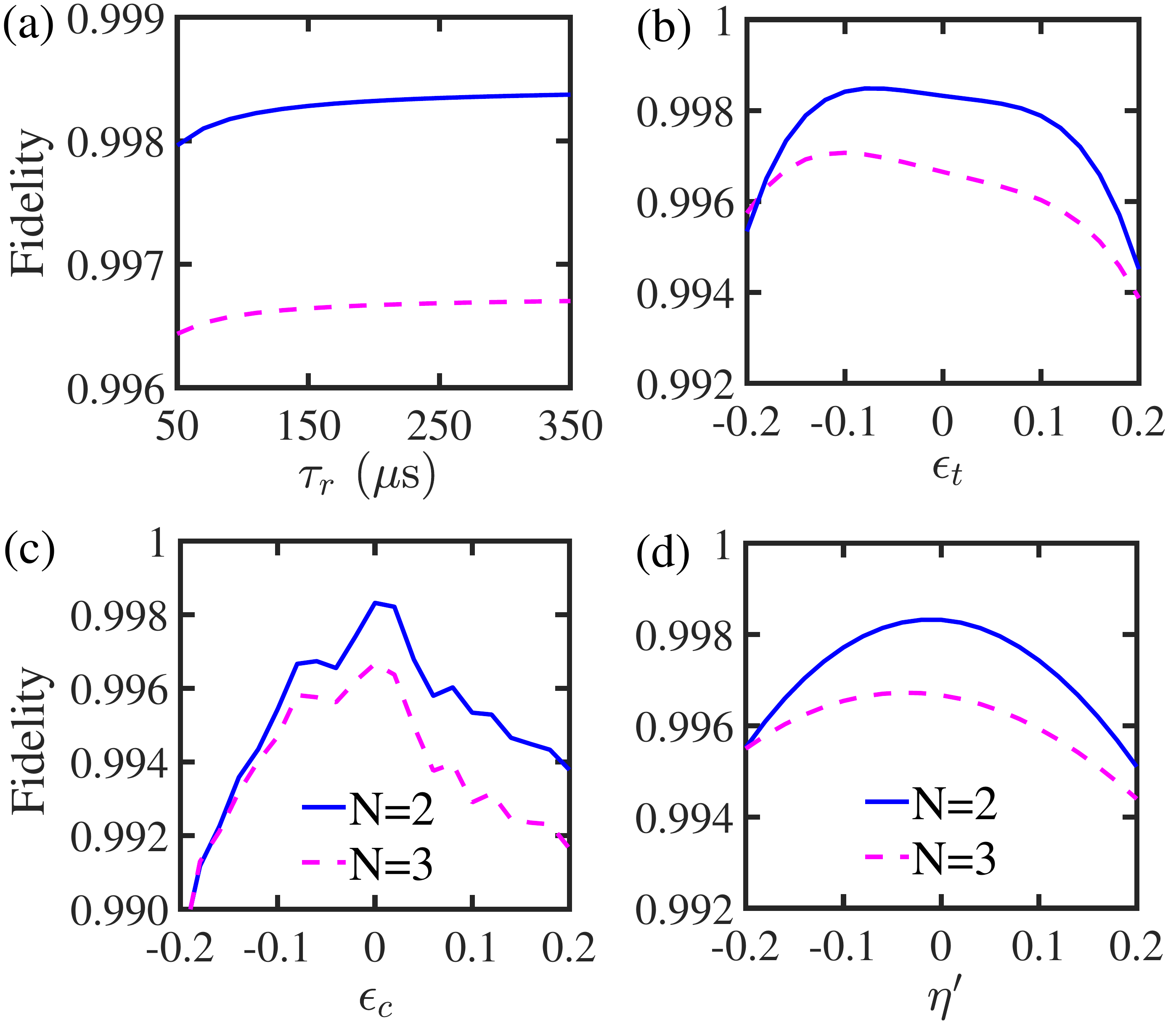}
\caption{  The performance of the $C_{N}Z$ gate implemented with the SINGQC scheme, where the solid blue line and the dashed purple line represent the results of the $C_2Z$ and $C_3Z$ gates, respectively. (a) Fidelities of the $C_{N}Z$ gate under different Rydberg state lifetimes. Fidelities of the $C_{N}Z$ gate with respect to the (b) control error of the target atom, (c) the control error of control atoms, and (d) the detuning error. }
\label{Fig7}
\end{figure}

\subsection{Gate performance}
Finally, we numerically test the performance of the SINGQC multiqubit gates that evolve along Path~1 by defining the gate fidelity as $F'=$  $\frac{1}{4^{N+1}}\sum_{j=1}^{4^{N+1}}\langle \Psi_j'(0)|U'^{\dag}\rho' U'|\Psi_j'(0)\rangle$, where $|\Psi_j'(0)\rangle=\otimes_i^{N+1}|\psi_i'(0)\rangle$ represents one of the initial states of the $N+1$ atom system and $|\psi_i'(0)\rangle$ denotes the $i$th atom initially in one of the states $\{|0\rangle,|1\rangle,(|0\rangle+|1\rangle)/\sqrt{2}, (|0\rangle-i|1\rangle)/\sqrt{2}\}$ \cite{YXu2020}. Here $U'$ is the evolution operator, and $\rho'$ is the density matrix of the multiqubit quantum system under consideration.
 Here we choose $\bar{\Omega}_c=2\pi \times 36$ MHz, which can be obtained experimentally by using a higher power in the blue beam and increasing the detuning from the intermediate level \cite{ Miroshnychenko2010,Huber2011,Ripka2018}. 
Other parameters are $ \Omega' =2\pi \times 0.75$ MHz, $V_t=\omega= 2\pi \times 400$ MHz, and $V_c= V_t/7$. The decoherence operators of the $k$ atom are $\sigma_{k}^{0}= |0\rangle_k\langle r|$, $\sigma_{k}^{1}=|1\rangle_k\langle r|$ and $\sigma_{k}^{2}= |2\rangle_k\langle r|$, where $|2\rangle$ is an additional ground state
representing the remainder of the Zeeman magnetic sublevels out of the computational states $|0\rangle$ and $|1\rangle$. For simplicity, we suppose that the decay rates of the Rydberg state to eight Zeeman ground states are the same. Thus, the decoherence rates are $\Gamma_{k}^0=\Gamma_{k}^1=\Gamma/8$, and $\Gamma_{k}^2=3\Gamma/4$, where $\Gamma=1/\tau_r$ with $\tau_r$ being the Rydberg state lifetime. 
Figure. \ref{Fig6}(a) plots the fidelities of the $CZ$ gate as a function of Rydberg state lifetime, where we find that the $CZ$ gate constructed in the SINGQC manner is more resistant to the finite Rydberg state lifetime than the typical Rydberg dynamical $CZ$ gate \cite{Jaksch2000}.  The fidelity of the SINGQC $CZ$ gate still exceeds $99.8\%$ even if the Rydberg state lifetime is only $50 \ {\rm \mu s}$. 
In the presence of systematic errors, including the control error $\epsilon_t$ ($\epsilon_c$) of the target atom (control atoms), and the detuning error $\eta'$, the SINGQC scheme is much more robust than the DG gate within the error range considered, as shown in Figs. \ref{Fig6}(b)-(d). In particular, the gate fidelity of our scheme can exceed $99.9\%$ when the fraction of $\epsilon_t$ or $\eta$ is within $5\%$. The Rydberg state lifetime is chosen to be $\tau_r= 200$ $\mu s$ \cite{Archimi2019,Barredo2020} in Figs. \ref{Fig6}(b)-(d).

Moreover, we also examine the robustness of $C_NZ$ ($N=2,3$) gates. As shown in Fig. \ref{Fig7}(a), despite the increase of atomic number, the SINGQC $C_{N } Z$ gates are extremely insensitive to Rydberg state lifetime. Even with a Rydberg state lifetime of $\tau_r= 50$ $\mu s$, the fidelity of SINGQC $C_{N } Z$ gates can exceed $99.6\%$. The fidelities as a function of the control error of the target atom are plotted in Fig. \ref{Fig7}(b), where the gate fidelities of both $C_2Z$ and $C_3 Z$ gates can be more than $99.3\%$ within the error rate range of $\epsilon_t\in[-0.2,0.2]$. In addition, the gate robustness to the control error of control atoms is shown in Fig. \ref{Fig7}(c), and the fidelities of $C_{N } Z$ gates can almost exceed the fault-tolerance threshold of the multiqubit quantum gate, i.e., $99\%$, within the error range considered. With regard to the robustness to detuning error, the SINGQC $C_{N } Z$ gates exhibit superior robustness with a fidelity exceeding $99.4\%$ within the considered detuning error range of $\eta'\in[-0.2,0.2]$, as shown in Fig. \ref{Fig7}(d). The Rydberg state lifetime is chosen to be $\tau_r= 200$ $\mu s$ in Figs. \ref{Fig7}(b)-(d).

\section{DISCUSSION AND CONCLUSION}
In conclusion, based on the inverse engineering of the Hamiltonian, we propose the SINGQC scheme, where arbitrary input states accumulate only geometric phases, which is different from the previous NGQC schemes. Numerical results indicate that our scheme can significantly improve the gate robustness against control error, and it can also enhance robustness against detuning errors through an alternative evolution path. 
In particular, the gate robustness of our scheme can outperform the DG scheme even when the rotation angle of the geometric gate is $\pi$, which breaks the limitation that the gate robustness of geometric schemes cannot exceed the DG scheme for the rotation angle of $\pi$ as was the case in previous schemes.

In addition, we construct the SINGQC multiqubit gates in the Rydberg atom system, where the gate operation time does not increase with the increase of the involved atom number. Numerical simulations show that the $CZ$ gate of our protocol is more robust than the DG scheme. Even for $C_2Z$ and $C_3Z$ gates, the gate fidelities of our scheme almost entirely exceed the fault tolerance threshold of the multiqubit gate within the considered error range. Moreover, our SINGQC scheme can also be applied to other solid-state platforms \cite{jianzhou, cxzhang, YXu2020}.

\begin{acknowledgements}
This work was supported by the National Natural Science Foundation of China (Grant No. 12275090) and the Guangdong Provincial Key Laboratory (Grant No. 2020B1212060066).
\end{acknowledgements}

\appendix

\section{ The dynamical scheme} \label{appendixA}
  In a generic two-level model, the dynamical gate is constructed by a simple resonant pulse, so the Hamiltonian of the system is
\begin{eqnarray}
\label{A4}
  H_d=\Omega e^{-i\varphi_d}|0\rangle\langle1|+{\rm H.c.}
\end{eqnarray}
The corresponding evolution operator can be expressed as
\begin{eqnarray}
\label{A5}
 U_d(\Theta_d,\varphi_d)=\cos\Theta_d \bm{I}-i\sin\Theta_d(\cos\varphi_d\sigma_x+\sin\varphi_d\sigma_y), \notag \\
\end{eqnarray}
where $\Theta_d=\Omega t$.  
%where $\tau$ is the duration time. Since an arbitrary single-qubit gate can be implemented by rotations around different unparallel axes in the x-y plane, we can easily obtain 
The $S$ gate and $H$ gate can be implemented as
\begin{subequations}
\label{A6}
\begin{align}
&U_d^S(\Theta_d,\varphi_d)=U_d(\frac{\pi}{4},\pi)U_d(\frac{\pi}{4},\frac{3\pi}{2})U_d(\frac{\pi}{4},0),\\   
&U_d^H(\Theta_d,\varphi_d)=U_d(\frac{\pi}{4},\frac{3\pi}{2})U_d(\frac{\pi}{2},0).
\end{align}
\end{subequations}
In the presence of errors, the Hamiltonian in Eq.(\ref{A4}) becomes $H'_d=(1+\epsilon)H_d+\eta\Omega\sigma_z/2$.

\section{ The conventional single-loop NGQC scheme} \label{appendix B}
 In single-loop NGQC scheme, the Hamiltonian in each segment is set to be \cite{TChen2018}
\begin{eqnarray}
\label{A7}
  H_s=\Omega e^{-i\varphi_s}|0\rangle\langle1|+{\rm H.c.}
\end{eqnarray}
The corresponding evolution operator is similar to that of Eq.(\ref{A5}), i.e., $ U_s(\Theta_s,\varphi_s)=\cos\Theta_s \bm{I}-i\sin\Theta_s(\cos\varphi_s\sigma_x+\sin\varphi_s\sigma_y)$ with $\Theta_s=\Omega\tau$. 
For the $S$ gate, the implementation is divided into two segments, that is,
\begin{eqnarray}
\label{A8}
  U_s^S(\Theta_s,\varphi_s)=U_s(\frac{\pi}{2},\frac{3\pi}{4})U_s(\frac{\pi}{2},-\frac{\pi}{2}).
\end{eqnarray}
 For the $H$ gate, three segments are needed and the evolution operator is 
\begin{eqnarray}
\label{A8}
  U_s^H(\Theta_s,\varphi_s)=U_s(\frac{3\pi}{8},-\frac{\pi}{2})U_s(\frac{\pi}{2},\pi)U_s(\frac{\pi}{8},-\frac{\pi}{2}).
\end{eqnarray}
Considering both types of error, the Hamiltonian becomes $H'_s=(1+\epsilon)H_s+\eta\Omega\sigma_z/2$.

\section{ The robustness to phase error}  \label{appendixC}

\begin{figure}[tbp]
  \centering
  \includegraphics[width=\linewidth]{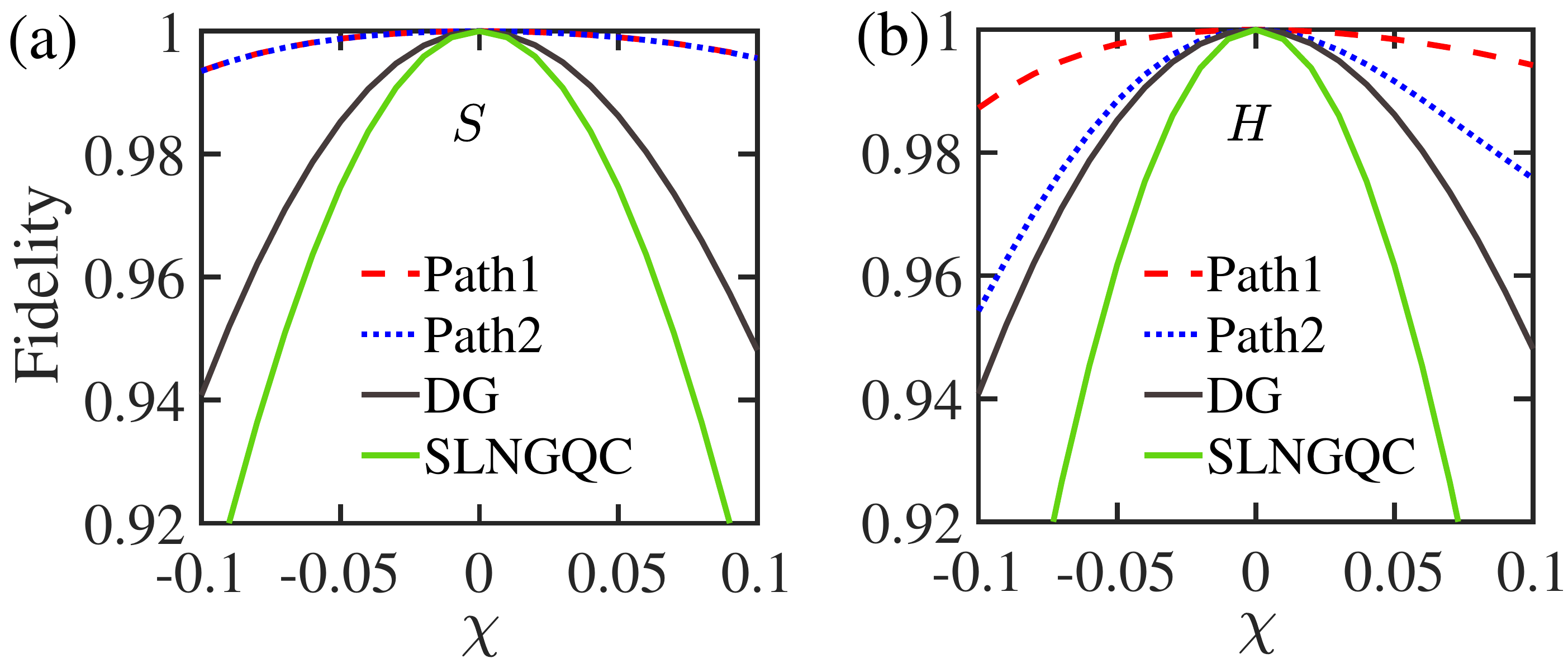}
\caption{ Gate fidelity as function of the phase error $\chi$ in the absence of decoherence. The results of the $S$ and $H$  gates are shown in (a) and (b), respectively, which indicate that our scheme 
is more robust against phase error than both the dynamical and single-loop NGQC schemes.} 
\label{Fig8}
\end{figure}

 To comprehensively show the performance of our scheme, we examine the robustness of our scheme against the phase error, i.e., $\Omega e^{i\phi}\rightarrow \Omega e^{i(1+\chi)\phi}$, where $\chi$ is the ratio of the phase error. In figure. \ref{Fig8}, we show the simulated gate fidelity as a function of $\chi$, in which we can see that our scheme is still more robust than the dynamical and single-loop NGQC schemes.

\end{document}